 \documentclass[final,5p,times,twocolumn,authoryear]{elsarticle}
\usepackage{amssymb}
\usepackage{lipsum}
\usepackage{amsmath} 
\usepackage{array, multirow}
\usepackage{graphicx}
\usepackage{upgreek}

\begin{document}

\begin{frontmatter}



\title{Ultra-low-energy and high-speed micro-laser-induced breakdown spectroscopy using GHz repetition rate pulses}


\author[1]{Ayesha Noor\fnmark[*]}
\affiliation[1]{organization={ Department of Electrical and Electronics Engineering},
            addressline={Özyeğin University}, 
            city={Istanbul},
            postcode={34794}, 
            country={Turkey}}
          
\author[2]{Emre Hasar\fnmark[*]}
\affiliation[2]{organization={ Department of Physics},
            addressline={Bogazici University}, 
            city={Istanbul},
            postcode={34794}, 
            country={Turkey}}
\author[3]{Parviz Elahi\corref{cor1}}
\affiliation[3]{organization={Department of Natural and Mathematical Sciences},
            addressline={Özyeğin University}, 
            city={Istanbul},
            postcode={34794}, 
            country={Turkey}
            }  

\cortext[cor1]{Corresponding author, Email: parviz.elahi@ozyegin.edu.tr}
\fntext[fn*]{Contributed equally}
\begin{abstract}
Using GHz repetition rate pulses in burst mode has demonstrated numerous advantages for highly efficient material removal. In this work, we present high-speed (over 100 kHz) micro-laser-induced breakdown spectroscopy (micro-LIBS) with ultra-low pulse energies in the range of 10-200 nJ using our recently developed 2.8 GHz burst-mode Yb-doped fiber laser. We delivered $\sim$40 ps-long pulses to the sample with a beam diameter of about 18 µm.
A systematic LIBS study was conducted on stainless steel (SS) at different burst durations and burst energies to study their effects on the optical emission spectrum. Finally, the electron temperature and electron density were determined using the Boltzmann plot method and Stark-broadened line profile analysis, respectively.
\end{abstract}
 


\begin{keyword}
GHz fiber laser \sep Burst mode fiber laser \sep Micro-LIBS \sep nJ LIBS \sep High-speed LIBS \sep Plasma temperature \sep Electron density



\end{keyword}

\end{frontmatter}




\section{Introduction}
\label{introduction}
During the past two decades, the laser-induced breakdown spectroscopy (LIBS) technique has been used in many industrial systems to analyze the elemental composition, for example, in environmental monitoring \cite{1}, material science \cite{doi:10.1366/11-06574}, mining \cite{harhira2017advanced}, food analysis \cite{stefas2021laser}, in geology to identify pollutants \cite{mahmood2024detection} and the composition of various substances \cite{kumar2022chemical}. LIBS also serves in forensic investigations \cite{naozuka2019laser}, space missions \cite{escudero2008optical}, and biomedical research \cite{rehse2012laser}. In addition, it is useful in archaeology, agriculture \cite{ren2022libs}, and pharmaceuticals for quality assurance \cite{lal2005laser}, and to analyze chemical compositions. Briefly, in LIBS a short laser pulse is focused onto a sample, which leads to the ablation of the material at the focused spot, and plasma is formed. The excited species inside the plasma plume emit the radiations at distinct atomic wavelengths, which can be detected using a spectrometer to determine the sample's elemental composition. LIBS offer a significant advantage over many other methods because it requires little or no sample preparation prior to analysis and enables fast, real-time testing, making it suitable for many applications. Moreover, LIBS can be applied to a wide range of materials, including solids, liquids, gases, and aerosols \cite{diaz2020libs}, \cite{agrup1987application}.

Many research on LIBS has focused on employing solid-state nanosecond lasers, such as Nd:YAG, typically operate with pulse energies in milijoules (mJ) range and repetition rates measured in hertz (Hz) \cite{elsayed2012design}, \cite{khumaeni2024spectrochemical}, \cite{khalepha2024comparative}. Nanosecond lasers generate pulses with high energy that are effective for material ablation; however, they are not suitable for thin solid samples because their prolonged interaction time and thermal effects can result in excessive material removal or damage. Additionally, the high-energy results in a broad continuum emission, which can complicate spectral analysis and necessitate the use of a gated spectrometer. In recent years, fiber laser technology has emerged as a promising alternative as a laser source is LIBS. Replacing nanosecond pulses with femtosecond pulses offers several benefits, ultra fast pulses reduce the ablation threshold, and the plasma is produced at low energies. It delivers energy within an extremely brief timeframe, preventing material damage by avoiding thermal effects \cite{article}, \cite{Baudelet2006}. They also provide advantages such as ambient pressure and matrix independence, along with high spatial resolution \cite{russo2004laser}, \cite{zorba2011ultrafast}. Furthermore, for developing portable LIBS systems, fiber lasers are applicable as they are more compact, energy-efficient, stable, reliable, and cost-effective. Solid-state lasers, on the other hand, are typically bulky, expensive, and sensitive to vibrations, which makes them ineffective for portable LIBS applications. Therefore, it is essential to examine LIBS using fiber laser. 

Few studies have investigated the impact of fiber laser parameters on the LIBS systems. For instance: \cite{Baudelet:10} studied the use of a 2 µm thulium fiber laser operating at pulse duration 200 ns, pulse energy 100 µJ and repetition rate of 20 KHz for LIBS on copper. \cite{PARKER2015146} also performed LIBS on copper by fiber laser generating fs pulses at 2 MHz and have reported that Yb fiber oscillator achieves a fivefold reduction in the LIBS ablation threshold in comparison to Ti:sapphire system. \cite{Huang:12} employed 1030 nm fiber laser with pulse duration 750 fs and 10 µJ pulse energy for the LIBS of various materials including metal, semiconductor and glass. \cite{C0JA00228C} investigated the lower detection limit (1.1 mg/g) of magnesium in aluminum by utilizing a compact nanosecond fiber laser operating at 1064 nm wavelength with a peak power of 11 kW.   

Although these researchers have explored the impact of fiber laser parameters on the LIBS applications, to the best of our knowledge no studies demonstrated a LIBS system employing a fiber laser operating in burst mode, with gigahertz (GHz) intra-burst repetition rates and nanojoule (nJ)-level pulse energy. 

Fiber lasers in burst mode, especially at GHz-range intraburst repetition rates, have gained considerable attention in recent years \cite{kerse20163, lal2005laser, liu2020100} and are increasingly being adopted in industry, particularly for high-efficiency material processing and micromachining \cite{kalayciouglu2017high, audouard2023high, bonamis2020systematic}. Generating GHz bursts of ultrashort pulses has significantly improved laser micromachining quality, in particular, using picosecond or femtosecond laser pulses in GHz burst mode introduces a novel ablation regime known as ablation cooling, where material removal occurs with minimal heat accumulation, enhancing material removal efficiency and ablation energy threshold \cite{kerse2016ablation}.\\

In this study, we introduce LIBS in the ablation-cooled regime for the first time, to the best of our knowledge, and demonstrate high-speed micro-LIBS at nJ-level pulse energy using a burst-mode fiber laser with a GHz intraburst repetition rate. We employed our recently developed all-polarization-maintaining Yb-doped fiber laser system operating in burst mode with a 2.8 GHz intra-burst repetition rate \cite{hasar2024high}. The system produced 40 ps pulses before compression, and 650 fs
pulses after compression at a central wavelength 1040 nm, with a maximum average power of 14 W. LIBS experiments were performed on stainless steel (SS) using three different burst durations 83 ns, 120 ns, and 240 ns corresponding to 232, 336, and 672 intraburst pulses, respectively, at a burst repetition rate of 100 kHz. The pulse energy varied from 9 nJ to 196 nJ. We analyzed the influence of burst duration, burst, and pulse energies on the optical emission spectrum of the SS. The electron temperature was calculated using the Boltzmann plot method, and the electron density via Stark broadening was determined. These measurements were performed for all three burst durations to examine how laser parameters affect plasma production. Additionally, we evaluated whether the plasma achieved local thermal equilibrium (LTE).

\section{Experimental Setup}
\label{Experimental Setup}

\begin{figure}[h]
    \centering 
    \includegraphics[width=8 cm]{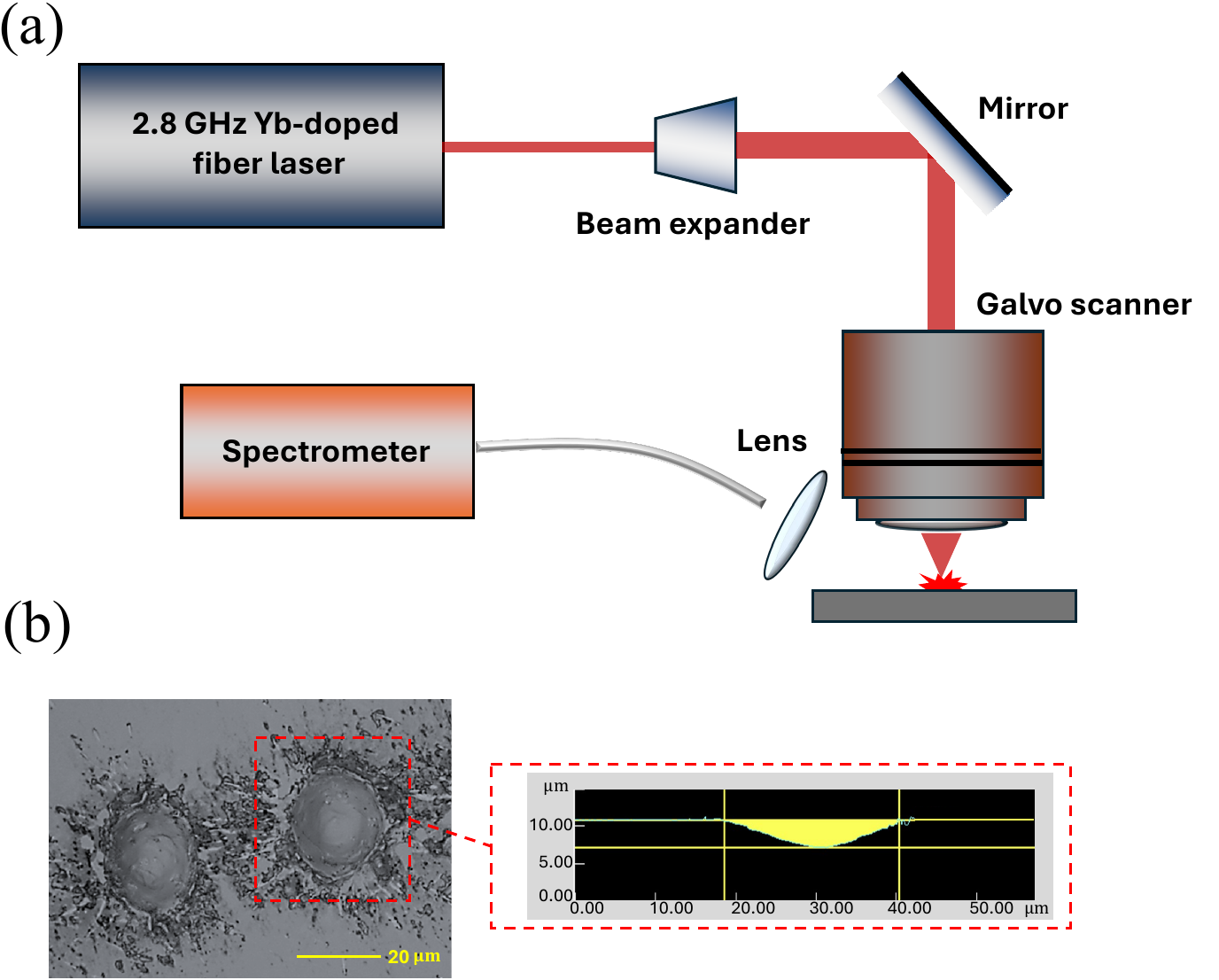}	
    \caption{(a) The schematic of the experimental setup. (b) An example of an ablated hole in SS and its 2D profile.} 
    \label{1}
\end{figure}

Figure \ref{1} illustrates the schematic representation of the experimental setup. We used our home-built high-power Yb-doped fiber laser operating in burst mode regime, with an intraburst repetition rate of 2.8 GHz and flexible burst repetition rate starting at 100 kHz \cite{hasar2024high}. The system produced 40 ps pulses before compression and 650 fs pulses after compression, with a maximum average power of 14 W. The output laser beam is directed into a Galvoscanner (Scanlab basiCube 10) which has a 10 mm entrance aperture and an f-theta lens with a focal length of 57 mm. The laser beam diameter was adjusted to approximately 8 mm with a beam expander, which has two converging lenses with focal lengths of 3.5 cm and 10 cm. 303 Stainless steel sample was placed on a 3D translation stage after the Galvoscanner to adjust a precise focus position. After determining the focus position, we employed the Galvoscanner with a square pattern at approximately 1.5 m/s in repetitive mode. Plasma was collected using a 2.5 cm focal length lens with the tip of the multi-mode fiber of the spectrometer (Ocean Optics USB4000) positioned precisely at the focal point of the lens. We performed a LIBS experiment on a steel sample with 3 different burst durations, which contained 232, 336, and 672 pulses. Due to the 30 ns rise and fall time of the acousto-optic modulator, we limited our burst width to 83 ns, which contains 232 pulses. To avoid amplified spontaneous emission, we set the burst repetition rate to be 100 kHz and greater \cite{gurel2014prediction} .            

\section{Result and Discussion}

\subsection{Analysis of Stainless Steel Spectrum and Spectral Features}
\begin{figure*}
\centering \includegraphics[width=18 cm ]{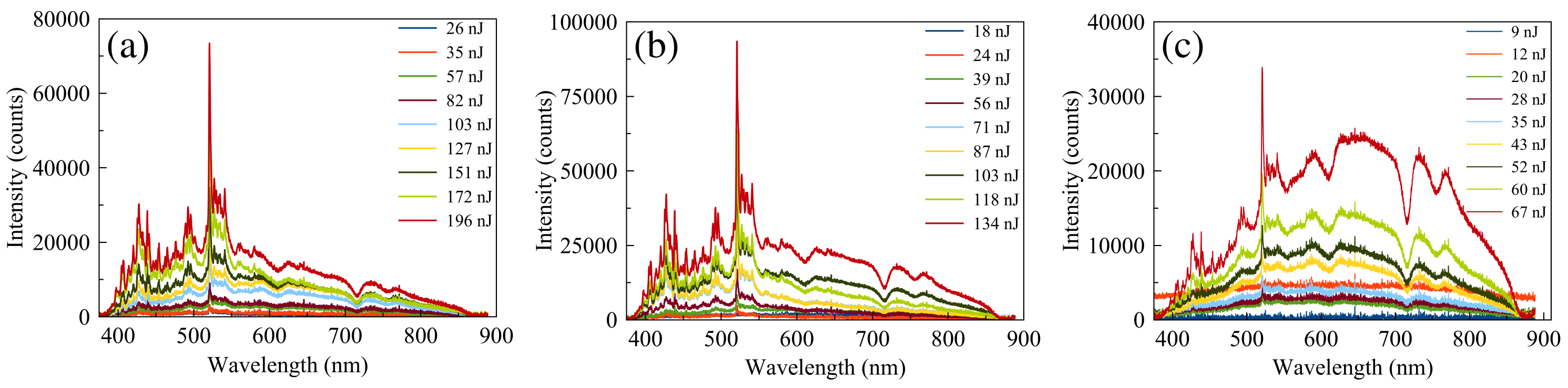}	
	\caption{Experimental measurement of LIBS spectra using 2.8 GHz pulse repetition rate laser at 100 kHz burst repetition rate on 303 stainless steel at different individual pulse energies at different intra-brust numbers of pulses: (a) 232 pulses/burst (b) 336 pulses/burst (c) 672 pulses/burst. The same color represents the same burst energy.} 
	\label{2}%
\end{figure*}
The optical spectrum recorded using a 2.8 GHz pulse repetition rate laser at 100 kHz burst repetition rate on SS at different intra-burst numbers of pulses at ambient air pressure, in a wavelength range of 200-900 nm is illustrated in Fig.\ref{2} (a-c). The influence of the laser pulse energy and burst duration on the spectrum of SS has been investigated. The effect of the laser pulse energy in a range of $ 26-196\ \text{nJ}$ corresponding to about 6-45 $\upmu\text{J}$/burst  on the emission spectra at 232 pulses/burst corresponding to 83 ns burst width is shown in Fig.\ref{2}(a).  it is visible that at low energies $ (26-82\ \text{nJ}) $, the emission lines are relatively weak and the signal-to-noise ratio (SNR) is low. This is attributed to insufficient material ablation, as the energy is close to the ablation threshold, and most of it is consumed in plasma formation. However, as the laser pulse energy increases, the intensity of these emission lines rises significantly. This is driven by more efficient material ablation and enhanced plasma excitation. At higher energies (above 103 nJ), line broadening becomes apparent, primarily due to Stark effects caused by increased electron density and ionization within the plasma. Additionally, the SNR is high and the continuum background intensifies because at high energies the plasma temperature rises and bremsstrahlung radiations become dominant \cite{harilal2014background}. A similar effect of laser pulse energy on the LIBS spectrum has been observed for the 336 and 672 pulses per burst (Figs. \ref{2}(b) and \ref{2}(c)). For these cases, the pulse energy ranges are $18-134\ \text{nJ}$ and $9-67\ \text{nJ}$, respectively corresponding to the same range of the burst energy as (a). Investigating the impact of intra-burst numbers of pulses in LIBS is essential, as it effects plasma behavior, signal strength and the precision of measurements.The intra-burst numbers of pulses indicate the total number of laser pulses produced within a single burst duration, during the operation of pulse laser in burst mode. It is given as, $ {N_\text{burst} = f_\text{rep}\times t_\text{burst}} $, where $f_\text{rep}$ is the repetition rate of the laser pulses within the burst and $t_\text{burst}$ is the burst duration. In this study, three intra-burst numbers of pulses were examined: 232, 336 and 672 pulses per burst. The case of 232 pulses per burst is significant for its shorter burst duration (83 ns), which minimizes thermal effects and consequently  reduces the continuum background effect. The sequential laser pulses in this setup help to sustain the plasma for a longer duration, enhancing the spectral line intensity and clarity of the signal. This improvement in signal quality is evident in the Fig.\ref{2}(a). With 336 pulses per burst, corresponding to a burst duration of 120 ns, the spectrum remains almost identical to the previous case, with only minor differences in the intensities of the emission lines. However, the thermal effect is slightly higher for 336 pulses, as shown in Fig.\ref{2}(b). When using 672 pulses per burst, corresponding to a 240 ns burst duration, the SNR is low while the thermal effects are very high . This is because the energy in this condition is too low to directly ablate the material. Instead, most of the energy is absorbed as heat by the target \cite{harilal2021spectro}. Subsequently, the intensity of the spectral lines is considerably lower compared to the pervious burst durations.
\begin{figure*}
	\centering 
\includegraphics[width=1\textwidth, angle=0]{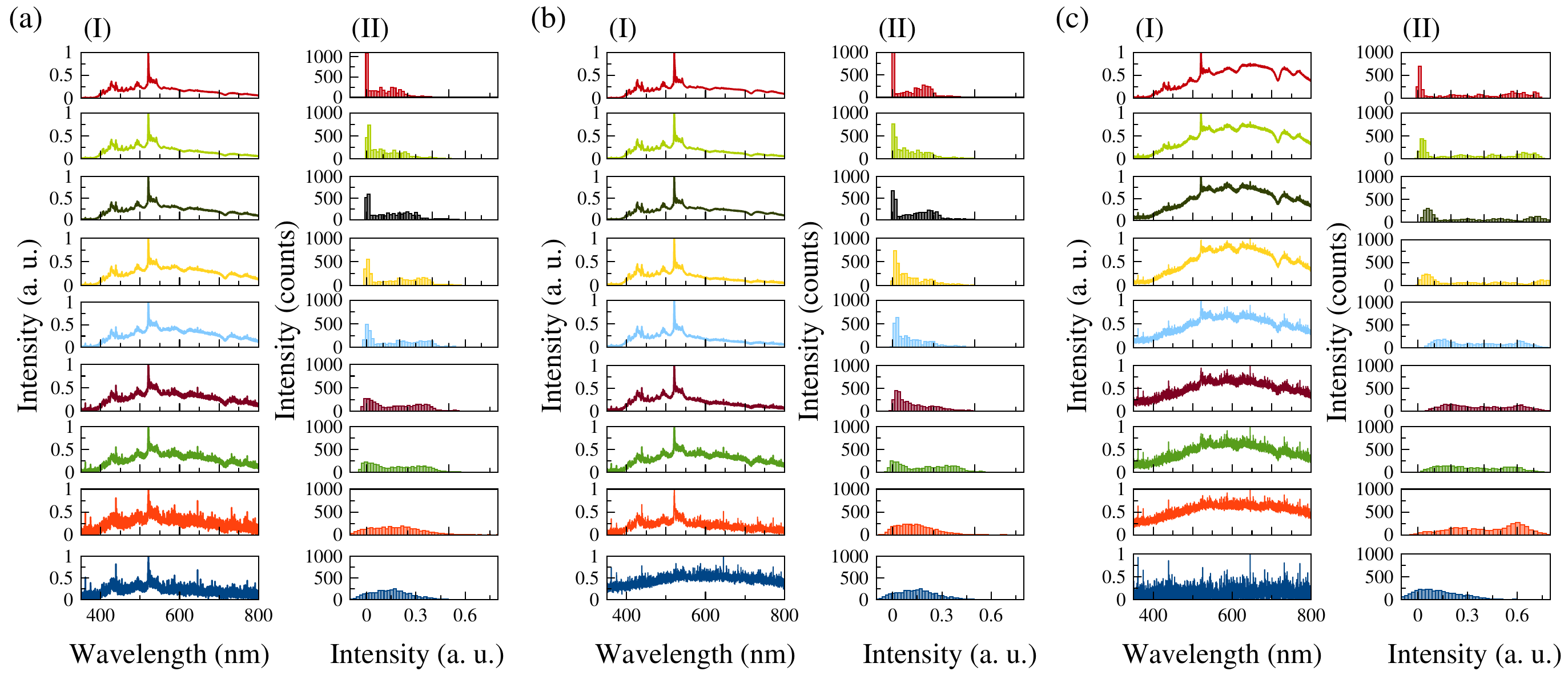}	
	\caption{(I) Normalized emission spectra of 303 stainless steel captured at different laser pulse energies, with (II) corresponding histograms illustrating the intensity distribution across each spectrum at different intra-brust numbers of pulses: (a) 232 pulses/burst (b) 336 pulses/burst (c) 672 pulses/burst. The pulse and burst energies are the same as Fig.\ref{2}. The sharper peak in the hostogram centered at zero indicates lower the blackbody effect in the spectrum.} 
	\label{3}%
\end{figure*}

In Fig.\ref{3}, the normalized emission spectra of SS are presented at different laser pulse energies, with corresponding histograms illustrating the intensity distribution across each spectrum at different intra-burst numbers of pulses. From Fig.\ref{3}(a-I) it is evident that for the burst duration 83 ns, at low energies (26 $\&$ 35 nJ) the emission lines are relatively weak but as the energy increases (above 57 nJ) beside the rise in continuum background the spectral lines are sharp and well defined. The corresponding intensity distribution in Fig.\ref{3}(a-II) is uniform indicating that at very low energy, specifically at 26 nJ, the plasma behaves like a blackbody radiator, emitting a continuous spectrum dominated by the thermal radiation. However, as the energy rises (above 57 nJ), the plasma reaches a non-equilibrium state with higher temperature and density, leading to increased ionization and emissions. Hence, the plasma no longer behaves as a blackbody radiator. This transition is clearly observed in the emission spectrum and the corresponding intensity distribution, which evolves to consist of sharp atomic and ionic lines replacing the smooth blackbody radiation spectrum. For a burst duration of 120 ns corresponding to 336 pulses, at pulse energy of 18 nJ, there is a complete absence of emission lines, as shown in Fig.\ref{3}(b-I).This indicates that the energy is too low, and thus, all the energy is absorbed by the target, leading to heating and producing a continuous spectrum. The intensity distribution in the Fig.\ref{3}(b-II) further demonstrate that this continuous emission spectrum is observed at first two energy levels. On the other hand, as the energy exceeds 39 nJ, the emission shifts from a broad spectrum to line emission. In the final case where the burst duration is 240 ns, the influence of pulse energy on the LIBS spectrum is clearly evident in the Fig.\ref{3}(c-I). It is observed that for the first four pulse energies no distinct atomic or ionic line emissions are present. Instead, the spectrum appeared as a broad, continuous signal with increased noise, primarily due to the dominance of thermal effect. Additionally, as the pulse energy increased, there is a noticeable rise in the continuum background and line broadening, which caused the spectral lines to become less distinct. This is further illustrated by the intensity distribution in Fig.\ref{3}(c-II), which shows a broad, continuous distribution resembling a blackbody spectrum. Moreover, it can be seen that for the 240 ns burst, the intensity distribution exhibits a greater spread, while for burst durations 83 ns and 120 ns, the intensity distribution becomes more concentrated and narrows around specific elemental lines, which subsequently confirmed that blackbody-like spectrum is dominant at 240 ns burst width. We note that at the same average power, a shorter burst duration leads to higher individual pulse energy, resulting in stronger ablation and reduced thermal effects, producing a highly ionized plasma with strong emissions. When thermal effects are minimized, a spectrum with sharp and well-defined peaks appears, and the blackbody-like spectrum also diminishes at small burst durations.

\begin{figure}[!ht]
	\centering 
\includegraphics[width=0.4\textwidth, angle=0]{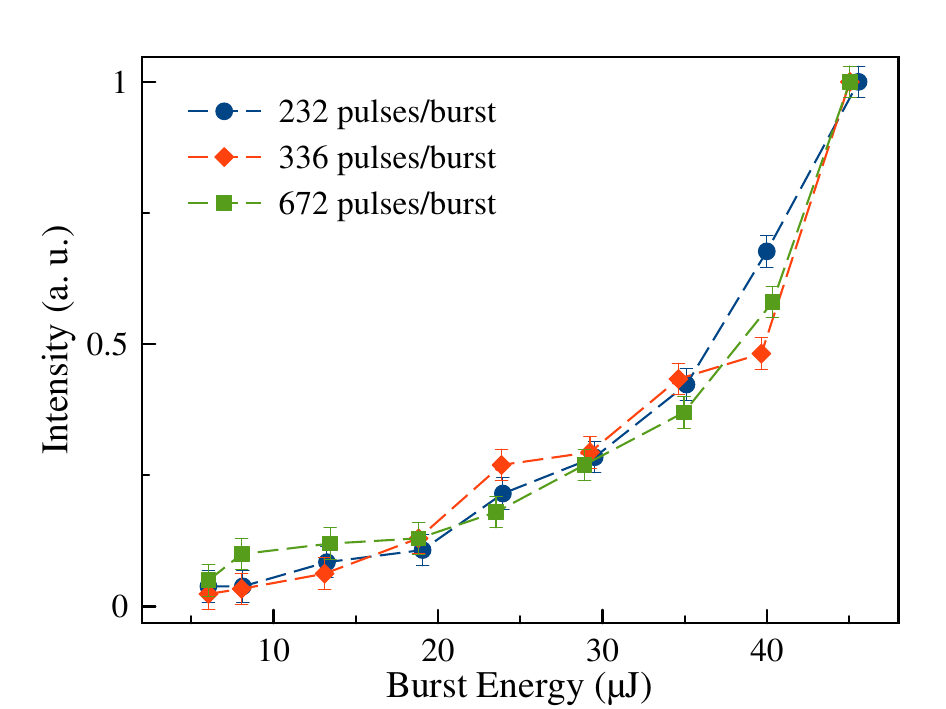}	
	\caption{The normalized intensity variation of Chromium (Cr I) peak at wavelength 520 nm versus burst energy for different intra-burst numbers of pulses.} 
	\label{4}
\end{figure}

Fig.\ref{4} depicts the relationship between the normalized intensity of chromium (Cr I) peak at a wavelength of 520 nm and burst energy. The data reveals a clear increasing trend in normalized intensity with the rise in burst energy, consistently observed across three distant intra-burst numbers of pulses: 232, 336, and 672 pulses per burst. This demonstrates a nonlinear dependence of the Cr I peak's optical intensity on burst energy.
However, it is crucial to note that while the trend remains consistent, this does not imply that the intensity counts are identical across different burst durations.

\subsection{Plasma Characterization: Electron Temperature and Density}
\begin{figure*}
	\centering 
\includegraphics[width=0.9\textwidth, angle=0]{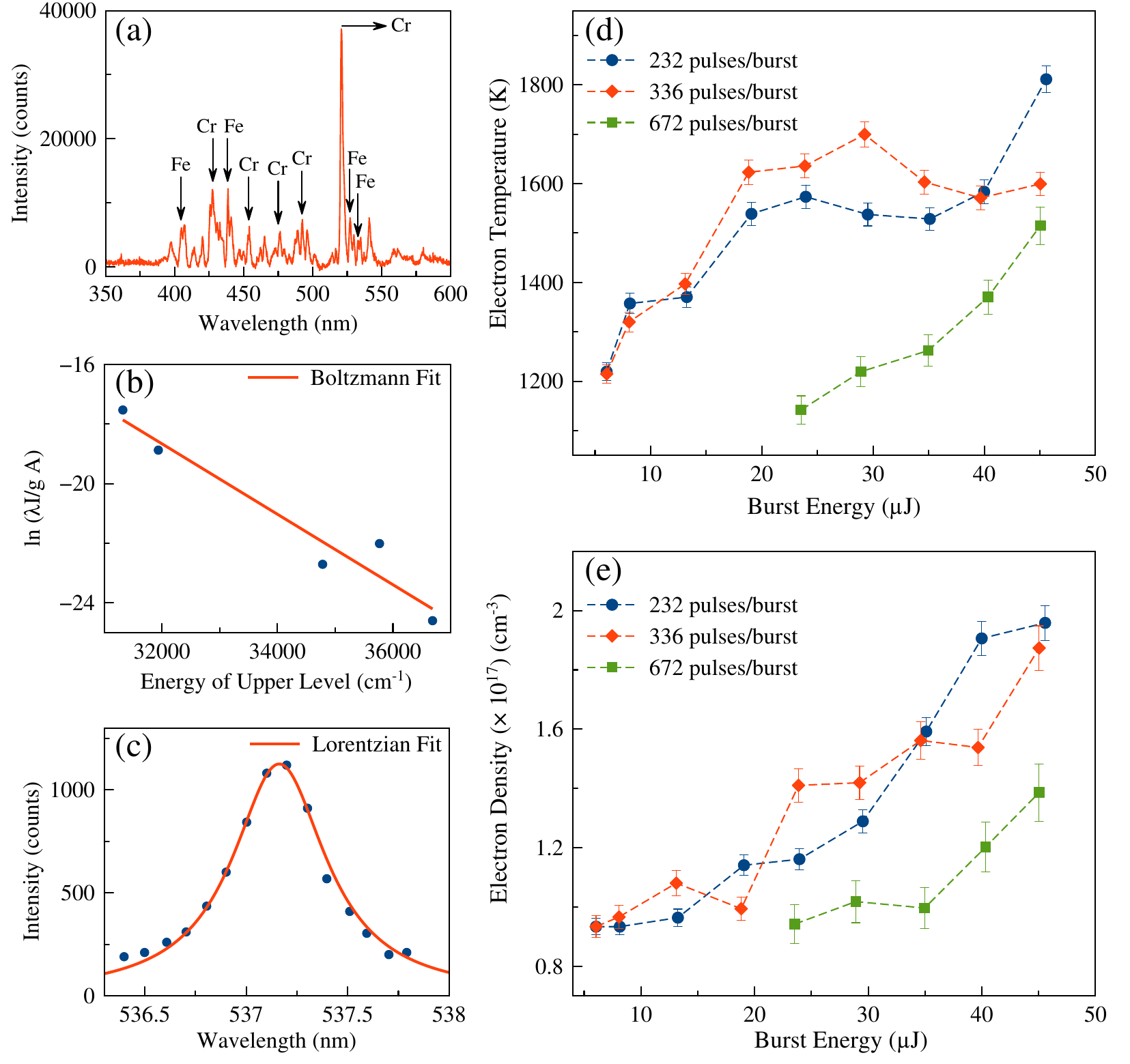}	
	\caption{(a) Baseline-corrected spectrum of SS showing the spectral lines of iron (Fe) and chromium (Cr) (b) Boltzmann plot of five spectral lines of Fe I, used to estimate the electron temperature (c) Stark- broadened line of Fe I, used to estimate the electron density. Variation of (d) electron temperature and (e) electron density with burst energy for different intra-brust number of pulses.} 
	\label{n_e and T_e}%
\end{figure*}
When a high-energy light source, such as a laser, interacts with the target, it induces processes such as ionization, excitations, and emissions. The electron temperature $ {T_\text{e}} $ refers to the average kinetic energy of the free electrons, and the electron density $ {n_\text{e}} $ is the number of free electrons within the plasma. Therefore, determining $ {T_\text{e}} $ and $ {n_\text{e}} $ in LIBS is essential to characterize the plasma properties.  Additionally, $ {T_{\text{e}}} $ and $ {n_{\text{e}}} $ directly depend on the laser parameters such as wavelength, pulse energy, pulse duration, and intra-burst duration. Understanding these relationships is crucial for optimizing the experimental conditions and enhancing the accuracy of LIBS measurements. It is important to note that the baseline correction was applied to all LIBS spectra of SS prior to the calculations of $ {T_\text{e}} $ and $ {n_\text{e}} $ to eliminate background effects.The corrected spectrum, shown in Fig.\ref{n_e and T_e}(a), highlights the characteristic emission lines of iron (Fe) at 404.58, 438.35, 527.03, 532.85 nm, along with the chromium (Cr) at 454.07, 473.73, 492.22, 520.60 nm. These distinctive lines confirm the spectral identification of the SS. The observed emission lines of iron and chromium primarily correspond to Fe I and Cr I, respectively. 
There are several methods for measuring the plasma temperature \cite{devia2015methods}, including the two-peak method \cite{unnikrishnan2010measurements}, continuum emission analysis \cite{Iordanova_2009}, and the Boltzmann plot method \cite{Guo:23}. Among these, the Boltzmann method is widely preferred for determining $ {T_\text{e}} $ due to its simplicity and reliability. Under the assumption that the emitted spectral lines are free from self-absorption and the plasma is in local thermal equilibrium (LTE) the Boltzmann equation is expressed as \cite{ARAGON2008893}, \cite{Guo:23}:

\begin{equation}   ln\left(\frac{I_{\text{mn}}\lambda_{\text{mn}}
    }{A_{\text{mn}}g_{\text{mn}}}\right)=ln\left(\frac{hcN}{4\pi Z}\right)-\frac{E_{\text{m}}
    }
    {kT_{\text{e}}}
\end{equation}
where, $ I_{\text{mn}} $ is the integrated spectral line intensity, $\lambda_{\text{mn}} $ is the wavelength, $  A_{\text{mn}} $ is the transition probability, $ g_{\text{mn}} $ is the statistical weight of the upper state, $E_{\text{m}} , k, $ and $ {T_{\text{e}}} $ are the energy of the upper state, Boltzmann constant and electron temperature, respectively. Using Equation (1) a grpah of $ln\left(\frac{I_{\text{mn}}\lambda_{\text{mn}}}{A_{\text{mn}}g_{\text{mn}}}\right) $ plotted against the $ E_{\text{m}} $ for the observed spectral lines results in a straight line as shown in Fig.\ref{n_e and T_e}(b). The slope of this line, given by $ -\frac{1}{kT_{\text{e}}} $ determines the electron temperature $ {T_{\text{e}}}$. In this study, five iron (Fe) spectral lines, specifically at 404.581, 420.209, 438.354, 527.035, 532.853 nm were selected to estimate $ {T_{\text{e}}} $. For the calculation of electron temperature the corresponding spectroscopic data for these transitions, obtained from the NIST data base \cite{nist_database} are listed in Table (1). 

\begin{table*}[!ht]
\centering
\renewcommand{\arraystretch}{2}
\scriptsize
\begin{tabular}{ c c c c c }
\hline
Wavelength (nm) & Transitions & Statistical Weight & Transition Probability (\(\text{s}^{-\mathbf{1}}\)) & Upper level Energy (\(\text{cm}^{-\mathbf{1}}\)) \\
\hline
        404.581  & $3d^{7}\ (^{4}F) \ 4s \rightarrow 3d^{7} \ (^{4}F) \ 4p$ & 4 & $ 8.62 \times 10^7 $ & 36686.176  \\
        420.209  & $3d^{7}\ (^{4}F) \ 4s \rightarrow 3d^{7} \ (^{4}F) \ 4p$ & 4 & $ 8.22 \times 10^6 $ & 35767.564  \\ 
        438.354  & $3d^{7}\ (^{4}F) \ 4s \rightarrow 3d^{7} \ (^{4}F) \ 4p$ & 4 & $ 5.00 \times 10^7 $ & 34782.421  \\ 
        527.035  & $3d^{7}\ (^{4}F) \ 4s \rightarrow 3d^{6} \ (^{5}D)\ 4s \ 4p \ (^{3}P^{o})$ & 2 & $ 3.67 \times 10^6 $ & 31937.325  \\ 
        532.853  & $3d^{6}\ (^{5}D) \ 4d \rightarrow 3d^{6} \ (^{5}D_{2}) \ 4f$ & 3 & $ 4.70 \times 10^5 $ & 31322.613  \\ \hline
    \end{tabular}
\caption{Spectroscopic parameters of Iron (Fe I) lines used for the estimation of electron temperature.}
\end{table*}
Another key parameter for plasma characterization is the electron density $ {n_{\text{e}}} $. In LIBS, the spectroscopic methods used for the estimation of $ {n_{\text{e}}} $ are Stark broadening \cite{pardini2013determination}, Saha-Boltzmann analysis \cite{unnikrishnan2010measurements} and spectral line intensity ratio. The spectral lines emitted by the laser-induced plasma exhibit significant broadening. This broadening arises from the Coulomb interactions between the emitting atoms with the free electrons and ions within the plasma, which allows the $ {n_{\text{e}}} $ to be determined from the line widths. Hence, the Stark broadening method is favored for $ {n_{\text{e}}} $ measurements because it directly relates the spectral line width to the $ {n_{\text{e}}} $. The electron number density is determined using the full width at half maximum (FWHM) of Stark broadened spectral lines and can be calculated with the following formula \cite{JTorres_2003} \cite{Zmerli_2010}: 
\begin{equation}   
\Delta \lambda_{1/2} = 2 \omega\left(\frac{N_e}{10^{16}}\right)+ 3.5 A\left(\frac{N_e}{10^{16}}\right)^{1/4}\times\left[ 1-{\frac{3}{4}} N_D^{-1/3}\right]\omega \left(\frac{N_e}{10^{16}}\right)
\end{equation}
where, $\Delta \lambda_{1/2}$ is the width of the spectral line, $\omega$ and $A$ are the electron and ion impact broadening parameters, respectively. $N_\text{e}$ are the electron density and $N_\text{D}$ the number of particles in the Debye sphere. This equation includes two parts: the first part accounts for the broadening of the line width due to the electron impact, while the second part reflects the broadening caused by the ion impact. The line width is mainly influenced by the first term because the effect of the ion impact broadening is minimal. As a result, the second term can be ignored, and the equation describing the spectral line width can be simplified as:
\begin{equation} 
\Delta \lambda_{1/2} = 2 \omega\left(\frac{N_e}{10^{16}}\right)
\end{equation}
The Fe I line $ 3d^7 (^4F) \ 4s \rightarrow 3d^6 \ ({}^5D) \ 4s \ 4p \ ({}^3P^\circ)$ at 537.14 nm has been used to determine the density of the electron number. The FWHM of the emission line is obtained by fitting the line profile with a Lorentzian function as presented in Fig\ref{n_e and T_e}(c), while the electron impact parameter $\omega$ is taken from the literature \cite{konjevic2002stark}. The actual line width $\Delta \lambda_{1/2}$ is calculated by removing the instrumental broadening from the measured line profile. By substituting the FWHM and $\omega$ in Eq. \ref{3}, the electron density is determined with an uncertainty of $10-15\%$.

After calculating the $ {T_{\text{e}}} $ and $ {n_{\text{e}}} $ for different intra-brust numbers of pulses, the results are plotted against the burst energy in Figs.\ref{n_e and T_e}(d) and (e). It is evident from the plot that the electron temperature and density rise with an increase in the burst energy. As the burst energy increases in a range ${(6-45)\space \ \text{µJ}} $, more energy is delivered to the target material, resulting in a higher energy density at the focal point \cite{PhysRevLett.88.097603}. This enhanced energy density, which causes greater photon-atom coupling, leading to stronger plasma formation. In hotter plasmas, more electrons are ejected from the atoms, increasing the frequency of collision processes (electron-electron and electron-ion). These collisions amplify the kinetic energy of the electrons, thereby raising the plasma temperature and density . At higher burst energies, increased collisions between electrons and ions lead to a rise in plasma frequency, resulting in plasma shielding. Consequently, the plasma does not absorb the incoming laser energy effectively as a result of this shielding effect, leading to the saturation of $ {T_{\text{e}}} $ at higher energy levels. The impact of another parameter, known as the number of pulses in the intra-burst region, on the $ {T_{\text{e}}} $ and $ {n_{\text{e}}} $ has also been investigated. The intra-brust pulse numbers considered are 232 pulses/burst, 336 pulses/burst, and 672 pulses/burst. Among these, the $ {T_{\text{e}}} $ and $ {n_{\text{e}}} $ are highest at 232 pulses/burst. This is attributed to the shorter burst duration and the extremely short interval between pulses, which prevents the dissipation of plasma generated by preceding pulses. Hence, the interaction between each new pulse and the partially ionized plasma from earlier pulses intensifies. Additionally, in this case the average laser power is high, which results in more material ablation, creating a dense plasma with increased collisions, this in turn leads to rise in the $ {T_{\text{e}}} $ and $ {n_{\text{e}}} $. For the case of 336 pulses/burst, there is a considerable overlap of pulses, and the average power is also elevated, which explains why the values of $ {T_{\text{e}}} $ and $ {n_{\text{e}}} $ are similar to those observed in the pervious case. However, at 672 pulses per burst, there is a noticeable decrease in the $ {T_{\text{e}}} $ and $ {n_{\text{e}}} $ values. This can be linked to the reduced average laser power in this case, which is close to the threshold, with the most energy being used to heat the target material. Additionally,due to the long intra-brust duration, the overlap between pulses is minimal, allowing the plasma formed by pervious pulses to cool down. The plasma shielding effect is also not significant in this scenario, as the incoming laser interacts directly with the target. Therefore, the electron density is lower, and with the plasma being less dense, fewer collisions occur, contributing to the reduction in the $ {T_{\text{e}}} $ in comparison to the other intra-brust numbers of pulses. 

For a plasma to be in local thermal equilibrium (LTE), the collision processes must dominate over the radiative processes. To verify the conditions necessary for LTE and establish the corresponding lower limit of ${n_{\text{e}}}$ the McWhirter criterion is applied, which is given as \cite{fujimoto1990validity}, \cite{CRISTOFORETTI201086}:
\begin{equation}
N_e \geq 1.6 \times 10^{12} T^{1/2} (\Delta E)^3
\end{equation}
where, $T$ is the plasma temperature (in kelvins), $\Delta E$ is represents the energy difference between the upper and lower energy levels (in eV). Using the Fe I spectral line at 527.035, with $\Delta E = 2.35\space\text{eV}$ and an average plasma temperature of $ {T_\text{e}} = 1502 \  \text{K}$, the criterion yields $N_e = 8.062\times 10^{14}\space \text{cm}^{-3}$. This value is significantly lower than the electron density obtained from the iron (Fe) line measurements. Hence, the electron density measured in this study exceeds the threshold required to satisfy the LTE condition.

\section{Summary and conclusions}
In this paper, we presented micro-LIBS in an ablation-cooled regime by sending bursts of nJ pulses at GHz repetition rate to the sample. The impact of burst duration and pulse energy on the optical emission spectrum of 303 stainless steel has been investigated. LIBS experiments were performed on SS using
three distinct burst durations: 83 ns, 120 ns, and 240 ns  with a burst repetition rate of 100 kHz. The experiments were conducted at various pulse energies ranging from 9 nJ to 196 nJ.  At a brust duration of 83 ns, corresponding to 232 pulses the optical spectrum shows a high SNR with strong signal intensity. The emission lines were sharp and the continuum background was minimal. In contrast, when using 672 pulses per burst with a 240 ns burst duration, significant thermal effects were observed, most  of the pulse energy is absorbed as heat by the target. Consequently, it was concluded that with a rise in burst duration there is a notable decrease in the intensity of the spectral lines and the appearance of a black body-like spectrum. Additionally, the intensities of Fe I spectral lines were employed to calculate the electron temperature, while the Stark broadened profile of the 537.14 Fe I line was analyzed to estimate the electron density. At a burst energy 45.5 µJ the electron temperature was measured as 1812 K, 1600 K and 1514 K for burst duration 83 ns, 120 ns and 240 ns respectively. Similarly, the corresponding electron densities were found to be $1.959\times 10^{17} $, $1.874\times10^{17} $, and $ 1.386\times10^{17} \text{cm}^{-3}$.

\section*{Acknowledgments}
This work is supported by TÜBITAK under projects numbered 121F392. PE thanks his former institute, Bogazici University.

\bibliographystyle{elsarticle-harv} 
\bibliography{nJLIBS}

\end{document}